\documentstyle[epsf,twocolumn]{jpsj}
\newcommand{\eqref}[1]{eq.~(\ref{eq:#1})}
\newcommand{\figref}[1]{Fig.~\ref{fig:#1}}
\newcommand{\tabref}[1]{Table \ref{tab:#1}}
\newcommand{\secref}[1]{\S~\ref{sec:#1}}

\newcommand{\vect}[1]{{\mbox{\boldmath $#1$}}}

\newcommand{\deffig}[3]{
  \begin{figure}[htbp]
  \begin{center}
    \leavevmode
    \epsfxsize=0.5\textwidth
    \epsfbox{#2}
    \caption{#3}
    \label{fig:#1}
  \end{center}
  \end{figure}
}
\def\runtitle{Kosterlitz-Thouless transition of quantum $XY$ model
  in two dimensions}

\def\runauthor{
Kenji {\sc Harada} and Naoki {\sc Kawashima}
}

\title
{Kosterlitz-Thouless transition of quantum $XY$ model in two dimensions}

\author
{ 
Kenji {\sc Harada} and Naoki {\sc Kawashima}$^1$
}

\inst
{
Department of Applied Analysis and Complex Dynamical Systems,\\
Graduate School of Informatics, Kyoto University,
Sakyo-ku, Kyoto 606-8501, Japan\\
$^1$
Department of Physics, Toho University, 
Miyama 2-2-1, Funabashi, Chiba 274-8510, Japan}

\recdate
{March 2, 1998}

\abst
{
  The two-dimensional $S=1/2$ $XY$ model is investigated
  with an extensive quantum Monte Carlo simulation.  The
  helicity modulus is precisely estimated through a
  continuous-time loop algorithm for systems up to
  $128 \times 128$ near and below the critical temperature.
  The critical temperature is estimated as $T_{\rm KT} = 0.3427(2)J$.
  The obtained estimates for the helicity modulus
  are well fitted by a scaling form
  derived from the Kosterlitz renormalization group
  equation.  The validity of the Kosterlitz-Thouless theory
  for this model is confirmed.
}

\kword
{
Quantum $XY$ model, Kosterlitz--Thouless transition, quantum Monte Carlo,
helicity modulus
}

\begin{document}
\sloppy \maketitle
%
\section{Introduction}
\label{sec:intro-XY}

The $XY$ model in two dimensions has been discussed in
various contexts such as magnets with easy-plane anisotropy,
superconductivity in a thin layer, granular superconducting
materials, and insulator-superfluid transition in He${}^4$
systems.  Naturally, a number of works were devoted to
clarifying the nature of the phase transition and the
low-temperature phase of this model.  Among them notable was
a large scale Monte Carlo simulation by Ding and
Makivi\'c\cite{DingM,MD}.  Based on computation of the
linear in-plane susceptibility and the correlation length at
various temperatures, they concluded that a phase transition
takes place at $T_{\rm KT} = 0.350(4)$\cite{DingM,MD} or
$0.353(3)$\cite{Ding}, and that this transition is of
Kosterlitz-Thouless~(KT)\cite{KT} type.  However, it is
technically difficult to distinguish an exponential
divergence from an algebraic one.  Because of this
difficulty, the validity of their conclusion on the nature
of the phase transition was
questioned\cite{Comment,CommentB}.  For the same reason, in
spite of their very extensive Monte Carlo calculation, Gupta
and Baillie\cite{GuptaB,GuptaC} did not draw a
definitive conclusion although their numerical results
seemed to suggest that the phase transition of the classical
$XY$ model is exactly what the KT theory predicts.

It should be noted that the above-mentioned technical
difficulty is partially due to the absence of
finite-size-scaling type analysis which is a common and
powerful tool in most numerical studies.  However, Sol\'yom and
Ziman\cite{Solyom} pointed out that the straightforward
application of the ordinary finite-size-scaling analysis
leads to a wrong result not only when the power-law
temperature dependence of the correlation length, which is
wrong, is assumed but also when the correct exponential
divergence is assumed.  They studied the size dependence of
the first excitation gap in the $S=1/2$ anisotropic $XXZ$
model in one dimension, which is exactly solvable and known
to have a transition of the KT type at the antiferromagnetic
isotropic point.  They found that the exact estimates for
the finite systems do not fit into the standard form of the
finite-size-scaling at the critical point.

Therefore, to obtain a definitive answer to the question
concerning the nature of the phase transition, we need to
have a correct form for the system size dependence of
quantities of interest.  In previous studies, to our
knowledge, a systematic study of such system size dependence
has been missing for the quantum $XY$ model.  There are,
however, some reports on classical $XY$ models.  Instead of
using the ordinary finite-size-scaling form, Weber and
Minnhagen\cite{WeberM} used the Kosterlitz renormalization
group equation\cite{Kosterlitz} for the data analysis in
their study of the classical $XY$ model in two dimensions.
They verified the KT type phase transition by comparing the
size dependence of the helicity modulus, $\Upsilon$, at the
critical temperature with the renormalization group flow
along the critical line that converges to the KT fixed point
with logarithmically slow convergence.  They observed not
only that the computed helicity modulus exhibits
the logarithmic dependence on the length scale but 
also that even the pre-factor of this logarithmic term agreed 
with the predicted one.  Following the same idea, Olsson\cite{Olsson}
performed a more detailed analysis of the classical model
with an extensive Monte Carlo simulation.  He observed that
the system-size dependence of the helicity modulus agreed
with the form derived from the Kosterlitz renormalization
group equation below and above the critical temperature as
well as right at the critical temperature.

The helicity modulus is known to exhibit the universal jump
at the critical temperature\cite{NelsonK}.  This quantity
corresponds to the super-fluid density when the model is
regarded as a Boson system with hard-cores.  In the
world-line quantum Monte Carlo method\cite{Suzuki}, the
helicity modulus is represented as the fluctuation in the
total winding number of world-lines by the following
equation\cite{Ceperley},
\begin{equation}
  \Upsilon = (T/2) \langle {\bf W}^2 \rangle,
  \label{eq:helicity}
\end{equation}
where ${\bf W} \equiv (W_x,W_y)$ with $W_x$ ($W_y$) being
the total winding number in the $x$ ($y$) direction.  It is
difficult to measure this quantity by means of a
conventional world-line quantum Monte Carlo method because a
conventional algorithm is not ergodic in that the winding
number is not allowed to vary.  In principle, it is possible
to make it ergodic by introducing additional global
movements of world-lines.  In practice, however, such
additional and in most cases ``ad-hoc'' global movements are
seldom accepted and the resulting estimates of the helicity
modulus have large statistical errors.  Therefore,
Makivi\'c\cite{Makivic} divided the whole system into a
number of sub-systems and measured the winding number for
each sub-system which may vary.  It is not totally clear if
this alternative way of measurement gives the same answer as the
conventional one does.  Another difficulty of the
conventional Monte Carlo method is its long auto-correlation
time near and below the critical temperature.  These
difficulties limited the accuracy and the precision of the
Makivi\'c's estimates of the helicity modulus and narrowed
the accessible temperature range of simulation.

In the present paper, we report some results of the quantum
Monte Carlo simulation of the $S=1/2$ $XY$ model using the
loop algorithm\cite{LA,LoopAlgorithm,LA1} with both discrete and
continuous imaginary time
representations\cite{ContinuousTime}.  Similar results for
smaller systems with discrete imaginary time representation
have been reported elsewhere\cite{Note}.  The use of the
loop algorithm eliminates both of the above-mentioned
difficulties.  There the number of particles as well as the
winding number can vary.  At the same time, there are a
number of reports\cite{LoopAlgorithm,LA1} on the drastic
improvement in the auto-correlation time of the simulation
by loop algorithms.  We aim at a detailed and precise
comparison between the quantum $XY$ model and the theory by
Kosterlitz\cite{Kosterlitz} through an accurate estimation
of the thermal fluctuation in the winding number near and
below the critical temperature.  We show that such an
estimation allows us to examine a new scaling form different
from the ordinary finite-size-scaling.  In this scaling
form, the distance from the critical point, i.e., $K-K_{\rm
  KT}$ appears in the form of $(K-K_{\rm KT}) (\log
(L/L_0))^2$, in contrast to $(K-K_{\rm KT}) L^{y_T}$ in the
ordinary finite-size-scaling.  At the same time, the
quantity $x \equiv \langle (\pi/2){\bf W}^2 \rangle - 2$
scales as $x \log (L/L_0)$ rather than $x/L^{\omega}$ with
some exponent $\omega$.  This ``scaling'' form is consistent
with Olsson's fitting functions.

In \secref{LA-XY}, we describe loop algorithms on discrete
and continuous imaginary time. In \secref{QMC-XY}, the
definition of helicity modulus, its improved estimator and
details of simulations are described. In \secref{UJ-XY},
estimates of the helicity modulus are presented and we
summarize the results in \secref{XY-Conclusion}.

\section{Monte Carlo Algorithms}
\label{sec:LA-XY}
\subsection{World-line Monte Carlo method for quantum $XY$ model}
The $S=1/2$ quantum $XY$ model is defined by the following
Hamiltonian.
\begin{equation}
  \hat{H} = \sum_{\langle \vect{i}\vect{j} \rangle}
  \hat{H}_{\vect{i}\vect{j}} = - J \sum_{\langle
    \vect{i}\vect{j} \rangle} (\hat{S}_{\vect{i}}^x
  \hat{S}_{\vect{j}}^x + \hat{S}_{\vect{i}}^y
  \hat{S}_{\vect{j}}^y),
   \label{eq:Hamiltonian}
\end{equation}
where $\langle \vect{i}\vect{j} \rangle$ runs over all
nearest-neighbor pairs on a square lattice.  As for the spin
operators, we use the convention in which
$(\hat{S}_\vect{i}^{\mu})^2 = 1/4$ $(\mu = x,y,z)$. We will
take $J$ as the unit of the energy scale in what follows.

In world-line Monte Carlo methods for $d$ dimensional
quantum systems, we generate world-line configurations in
the $(d+1)$ dimensional space-time with probability density
proportional to the integrand of the Feynmann path integral.
In a conventional
quantum Monte Carlo simulation, it is customary to
discretize the imaginary time in the path integral form to
derive a model with classical degrees of freedom in $(d+1)$
dimensions.  We then perform an ordinary Metropolis-type
Monte Carlo simulation of this classical system.  This
standard procedure is called Suzuki--Trotter~(ST)
decomposition\cite{Suzuki}.  Due to the recent development
of the continuous time representation \cite{ContinuousTime},
it is no longer necessary to discretize the imaginary time
by the ST decomposition.  Nevertheless, we start with it
because it is still the most natural way to explain the
algorithm.  In the actual computation, we have used both of
the discrete and continuous time algorithms.

Our Hamiltonian can be considered as a sum of four
sub-Hamiltonians: $\hat H = \hat H_A+\hat H_B+\hat H_C+\hat H_D$.  Each
sub-Hamiltonian is a sum of operators commutable with each
other, i.e.,
\begin{equation}
  \hat{H}_X = \sum_{\langle\vect{i}\vect{j}\rangle\in X}
  \hat{H}_{\vect{i}\vect{j}} \quad (X=A,B,C,D)
  \label{eq:HamiltonianDecomposition}
\end{equation}
where
\begin{eqnarray}
  \label{eq:sets-four}
  A &\equiv& \{\langle \vect{i}\vect{j} \rangle | \vect{i}
  \in \mbox{odd column}, \vect{j}=\vect{i}+\vect{e}_x\},\\ 
  B &\equiv& \{\langle \vect{i}\vect{j} \rangle | \vect{i}
  \in \mbox{odd row}, \vect{j}=\vect{i}+\vect{e}_y\},\\
  C &\equiv& \{\langle \vect{i}\vect{j} \rangle | \vect{i}
  \in \mbox{even column}, \vect{j}=\vect{i}+\vect{e}_x\},\\
  D &\equiv& \{\langle \vect{i}\vect{j} \rangle | \vect{i}
  \in \mbox{even row}, \vect{j}=\vect{i}+\vect{e}_y\},
\end{eqnarray}
and $\vect{e}_{x}$ (or $\vect{e}_{y}$) is a unit lattice vector in
the x (or y) direction.

Using the ST decomposition, we transform the partition
function into the following form,
\begin{equation}
  Z \approx \sum_{S} \prod_{p}w(S_p)
  \label{eq:ClassicalRepresentation}.
\end{equation}
Here, $p$ stands for a plaquette in the $(d+1)$ dimensional
space-time with two edges perpendicular and the other two
parallel to the imaginary time axis.  The space-time
location of its left-bottom corner is given by
$(\vect{i},t)$ with
\begin{equation}
  \label{eq:cube-plaquette}
  t \equiv \left\{\begin{array}{lll} 
       (4n)\Delta\tau   & \mbox{if} &\vect{i} \in A\\ 
       (4n+1)\Delta\tau & \mbox{if} &\vect{i} \in B\\ 
       (4n+2)\Delta\tau & \mbox{if} &\vect{i} \in C\\
       (4n+3)\Delta\tau & \mbox{if} &\vect{i} \in D
  \end{array}\right.,
\end{equation}
where the imaginary time spacing, $\Delta\tau\equiv\beta/m$,
is the unit of the discretization of the imaginary time.
The number of steps $m$ is called the Trotter number.  
The symbol $S_p$ in \eqref{ClassicalRepresentation}
is the local spin configuration on a plaquette $p$ and
$w(S_p)$ is the two-spin propagator defined below.  The
symbol $S$ in \eqref{ClassicalRepresentation} stands for the
spin configuration of the whole space-time or the union of
all $S_p$'s, i.e., $S \equiv \bigcup_p S_p$.  We denote the
four states of spin pairs by $1 = \uparrow \uparrow$, $2
=\uparrow \downarrow$, $3 = \downarrow \uparrow$ and $4 =
\downarrow \downarrow$. Then, the two-spin propagator can be
written explicitly as
\begin{eqnarray}
  w(S_p) & \equiv & \left\langle S_p^{({\rm final})} \left| \exp
      \left( - (\Delta\tau) \hat{H}_{\vect{i}\vect{j}}
      \right) \right| S_p^{({\rm initial})} \right\rangle
  \nonumber \\ & = & \left(
    \begin{array}{cccc}
      1 & 0 & 0 & 0 \\ 0 & \cosh ( \frac{\Delta\tau J}{2} ) &
      \sinh ( \frac{\Delta\tau J}{2} ) & 0 \\ 0 & \sinh (
      \frac{\Delta\tau J}{2} ) & \cosh ( \frac{\Delta\tau J}{2} )
      & 0 \\ 0 & 0 & 0 & 1 \\ 
    \end{array}
  \right).
  \label{eq:tspm}  
\end{eqnarray}
The symbols $S_p^{({\rm initial})}$ and $S_p^{({\rm final})}$
stand for the local state of two corners at the bottom and
the top edge of the plaquette $p$, respectively.  The local
state of the whole plaquette, $S_p$, can be regarded as the
combination of the two.  In the $4\times 4$ matrix
representation of \eqref{tspm}, the column index corresponds
to four possible initial state, $S_p^{({\rm initial})}$, and
the row index $S_p^{({\rm final})}$.

\subsection{Loop algorithm with discrete imaginary time}
\label{sec:LA-XY-d}
In a loop algorithm, we first assign a graph $G_p$ to each
plaquette. A graph specifies the set of local states from
which new state $S'_p$ is chosen.  In other words, $G_p$
imposes a restriction on the local state.  We call it a
``graph'' because such a restriction can be conveniently
expressed pictorially by drawing a line connecting two sites
whose relative orientation is not allowed to be changed
under the restriction $G_p$.  A group of spins connected by
a sequence of such lines is called a cluster.  In the
present case each site is shared by two plaquettes and it is
connected to another site in each of the two plaquettes.
Therefore, a cluster is a loop in the present case.  Hence
the name of the algorithm.  In general, however, we need to
introduce branching or connecting more than two sites in a
plaquette.  For example, in the case of the Ising-like $XXZ$
model, we need to include a graph in which all four corners
are connected.  Once we assign graphs to all plaquettes, we
change the spin variables by regarding each cluster as a block
spin and flipping it with an appropriate probability.  In
what follows, we briefly describe how we should choose the
probability for assigning a graph to each plaquette and the
probability for changing the spin variables.

First, the two-spin propagator or the local Boltzmann weight
$w(S_p)$ can be written as
\begin{eqnarray}
\label{eq:w-decomp}
w(S_p) & = & \sum_{G_p} w(S_p,G_p), \nonumber \\ 
w(S_p,G_p) & \equiv &
v(G_p)\Delta(S_p,G_p). \label{eq:WeightEquation}
\end{eqnarray}
Here the function $\Delta(S_p,G_p)$ takes on the value 1 when
$S_p$ is compatible with the graph $G_p$, and it takes the
value 0, otherwise. 

For the $S=1/2$ quantum $XY$ model, the graph weight, $v(G_p)$,
is defined in \tabref{v-ed-xy}. The summation in
\eqref{WeightEquation} is taken over the three graphs
depicted in \figref{breakups-xxz}.  We represent
$\Delta(S_p,G_p)$ by the matrices in the same manner as
\eqref{tspm}:
\begin{eqnarray}
  \Delta(\cdot,1) &=& \left(\begin{array}{cccc} 1 & 0 & 0 &
      0\\ 0 & 1 & 0 & 0\\ 0 & 0 & 1 & 0\\ 0 & 0 & 0 & 1
  \end{array}\right),
\Delta(\cdot,2) = \left(\begin{array}{cccc} 0 & 0 & 0 & 0\\ 
    0 & 1 & 1 & 0\\ 0 & 1 & 1 & 0\\ 0 & 0 & 0 & 0
  \end{array}\right),\nonumber\\
\Delta(\cdot,3) &=& \left(\begin{array}{cccc} 1 & 0 & 0 &
    0\\ 0 & 0 & 1 & 0\\ 0 & 1 & 0 & 0\\ 0 & 0 & 0 & 1
  \end{array}\right).
\end{eqnarray}

A Markov process generated by a traditional world-line
algorithm stays in the spin configuration space:
\begin{equation}
\label{eq:mpot}
S^{(1)} \to S^{(2)} \to S^{(3)} \to \cdots,
\end{equation}
whereas in a Markov process of a loop algorithm the spin and the
graph configuration space alternate:
\begin{equation}
\label{eq:mpocl}
S^{(1)} \to G^{(1)} \to S^{(2)} \to G^{(2)} \to S^{(3)} \to
G^{(3)} \to \cdots.
\end{equation}
The transition probabilities between the spin and the graph
configuration are defined as follows:
\begin{eqnarray}
  \label{eq:stog}
  P(S \to G) & \equiv & \frac{\prod_p
    w(S_p,G_p)}{\sum_{G'_p} \prod_p w(S_p,G'_p)},\\ 
  \label{eq:gtos}
  P(G \to S) & \equiv & \frac{\prod_p
    w(S_p,G_p)}{\sum_{S'_p} \prod_p w(S'_p,G_p)},
\end{eqnarray}
where $G \equiv \bigcup_p G_p$ is the global graph.  It is
easy to check that these transition probabilities satisfy
the detailed balance.

We see from \eqref{gtos} that $P(G \to S)$ is vanishing if
$S_p$ is not compatible to $G_p$ for any $p$, while it takes on
a value independent of $S$ as long as all $S_p$'s are
compatible to $G_p$'s.  This means that sites belonging to
the same loop should be flipped simultaneously with a
probability 1/2, and two distinct loops should be flipped
independently.  Since flipping a loop is a global update,
one may imagine that the cluster algorithm reduces the
autocorrelation time due to slow relaxation modes associated
with large structures.  This is indeed the case as reported
in a number of articles\cite{LoopAlgorithm,LA1}.

\subsection{Loop algorithm with continuous imaginary time}
\label{sec:LA-XY-nd}
Beard and Wiese\cite{ContinuousTime} pointed out that we can
take the continuous imaginary time limit
($\Delta\tau\rightarrow 0$) of the above-mentioned algorithm
to obtain another algorithm which works directly on
continuous time, instead of computing quantities using
discrete time with various $\Delta\tau$'s and extrapolating
the results to get estimates in the continuous time limit.
In the discrete time version, we need to be careful in
decomposing the Boltzmann operator in order to keep the
correction term of high order in $\Delta\tau$.  This is why
we have to split the Hamiltonian into several pieces as we
did in previous sections (\eqref{HamiltonianDecomposition}).
In deriving the continuous time version, however, we can
avoid this complication because any correction term vanishes
in the continuous time limit.  In other words, we have to
keep only the lowest order term to get the correct
algorithm.  This fact makes the derivation of the algorithm
a little simpler as we see below.

The matrix elements of the partial Hamiltonian
$\hat{H}_{\vect{i}\vect{j}}$ can be directly represented by
a graph variable $G_p$ as
\begin{equation}
  \label{eq:ph}
  \langle S_p^{({\rm final})} | \hat{H}_{\vect{i}\vect{j}} |
  S_p^{({\rm initial})} \rangle \equiv - \sum_{G_p} a(G_p)
  \Delta(S_p,G_p).
\end{equation}
(For the $S=1/2$ quantum $XY$ model, the $a(G_p)$ is defined
in \tabref{v-ed-xy}).  Then, for small $\Delta\tau$, the
weights $w(S_p)$ are expressed, up to the lowest order, as
\begin{equation}
  \label{eq:w-ll}
  w(S_p)
  \approx I(S_p) + \Delta\tau \sum_{G_p} a(G_p)
  \Delta(S_p,G_p),
\end{equation}
where $I(S_p)$ is the matrix element of the identity
operator on a plaquette with a ``height'' $\Delta\tau$ in
the imaginary time direction. For simplicity, we set
$\Delta(S_p,1) \equiv I(S_p)$ in the following.  Taking the
limit of an infinitesimal imaginary time spacing $(\Delta\tau
\to 0)$, transition probabilities (\eqref{stog}) on a
plaquette are reduced to the followings:
\begin{enumerate}
\item If the present state $S_p$ is compatible to the
  graph $G_p = 1$, i.e., if there is no exchange of spins
  in the time interval $(t, t+\Delta\tau)$,
  \begin{eqnarray}
    & & P(S_p \to G_p) = (\Delta\tau) a(G_p) \Delta(S_p,G_p)
    \quad (G_p \ne 1), \nonumber \\ & & P(S_p \to 1 ) = 1 -
    \sum_{G'_p \ne 1} P(S_p \to G'_p).
    \label{eq:tpd}
  \end{eqnarray}
\item If the present state $S_p$ is incompatible to the
  graph $G_p = 1$, i.e., if the state at $t$ is different
  from that at $t+\Delta\tau$,
  \begin{eqnarray}
    & & P(S_p \to G_p) =
    \frac{a(G_p)\Delta(S_p,G_p)}{\sum_{G'_p \ne
        1}a(G'_p)\Delta(S_p,G'_p)} \quad (G_p \ne 1),
    \nonumber \\ & & P(S_p \to 1 ) = 0
    \label{eq:tpnd}
  \end{eqnarray}
\end{enumerate}

We apply \eqref{tpd} to plaquettes (with infinitesimal
``heights'') at which world-lines continue along the Trotter
direction.  The probability of choosing a graph $G_p \ne 1$
in the imaginary time interval $\Delta\tau$ is $(\Delta\tau)
a(G_p) \Delta(S_p,G_p)$. This means that in the continuous
limit we distribute graphs $G_p$ over such an interval
uniformly, i.e., with a Poisson process, with the
probability density $a(G_p)\Delta(S_p,G_p)$.  On the other
hand, \eqref{tpnd} gives the probability (not probability
density) with which a graph is assigned to each point of
time where the local state is changed or two neighboring
world-lines are exchanged.

Consequently, for general models, the loop algorithm with
continuous imaginary time can be summarized as follows.  For
each pair of nearest neighbor world-lines,
\begin{enumerate}
\item distribute graphs $G_p(\ne 1)$ with a Poisson
  process with \eqref{tpd} over every imaginary time interval
  in which world-lines are not interrupted,
\item choose a graph $G_p$ with \eqref{tpnd} at each point
  of time where states are exchanged between the two
  sites,
\item assign the graph $G_p = 1$ elsewhere,
\end{enumerate}
and then update spin values by flipping clusters
(\eqref{gtos}).  To restate this procedure for the specific
case of $S=1/2$ quantum $XY$ model,
\begin{enumerate}
\item for each uninterrupted time interval during which
  spins on these world-lines are antiparallel, generate
  ``horizontal reconnections'' (graph $G_p = 2$ in
  \figref{breakups-xxz}) of world-lines with probability
  density $J/4$, and for parallel spins, generate ``diagonal
  reconnections'' (graph $G_p = 3$) of world-lines with
  probability density $J/4$,
\item at each point of time where states are exchanged,
  assign a horizontal or diagonal reconnection with equal
  probability (1/2),
\item assign the graph $G_p = 1$ elsewhere.
\end{enumerate}

\section{Details of the Computation}
\label{sec:QMC-XY}
\subsection{Helicity modulus}
\label{sec:Helicity-def}
In the present paper, we focus on the helicity modulus
because it may be directly related to the solution of the
Kosterlitz renormalization group equation and therefore may
exhibit some behavior characteristic of the KT transition.
We have measured the helicity modulus by computing the
fluctuation in the total winding number of world-lines with
\eqref{helicity}.  The total winding number $W_x$ (or $W_y$)
is defined as the sum of winding numbers of individual
world-lines.  The winding number of an individual world-line
is the number of times the world-line wraps around the
system in the $x$ or $y$ direction before coming back to its
starting point.  The total winding number of world-lines for
up spins is exactly the same in magnitude as that for down
spins and has the opposite sign.  We here count only the
winding numbers for up spins.  Alternatively, $W_x$ can be
defined as
\begin{eqnarray}
  & & W_x = \frac{1}{L_x}\sum_p \alpha_x(S_p), \nonumber \\ 
  & & \frac{1}{L_x} \alpha_x(S_p) \equiv 
      \sum_{\vect{i} \in p} c_{\vect{i}}^x S_{\vect{i}}^z,
  \label{eq:Wx}
\end{eqnarray}
where $L_x$ is the lattice size in the $x$ direction.  The
symbol $\alpha_x(S_{p})$ stands for the function which takes on
the value $1$ (or $-1$) if an up-spin world-line passes
through the plaquette $p$ in the positive (or negative) $x$
direction and takes on the value 0, otherwise.  Equivalently, $c_\vect{i}^x$
is a constant which only depends on site $\vect{i}$ and takes on
the value $\pm 1/(4L_x)$ or zero.  The other winding number
$W_y$ can be also calculated in the same manner.

\subsection{Improved estimator of helicity modulus}
\label{sec:IE-Helicity}
In cluster type algorithms, it is often advantageous to
re-express physical quantities in terms of graph variables
rather than spin variables.  The best known example is the
graphical estimator for the linear magnetic susceptibility
in the Swendsen--Wang algorithm for the uniform Ising model
where the susceptibility is measured as the average cluster
size rather than  as the second moment of the sum of all spin
variables.  For the winding number, too, we can use a
similar improved estimator. In our simulation, we have used
this improved estimator for the squared winding number of
world-lines. Equation (\ref{eq:Wx}) defines the ordinary
estimator in terms of spin variables and it can be rewritten as
\begin{eqnarray}
  & & \vect{W}^2 = W_x^2+W_y^2, \\ 
  & & W_x \equiv \sum_{\vect{i}} c^x_{\vect{i}} S_{\vect{i}}^z, 
     \quad W_y \equiv \sum_{\vect{i}} c^y_{\vect{i}} S_{\vect{i}}^z,
\end{eqnarray}
where $c^x_{\vect{i}}$ and $c^y_{\vect{i}}$ are constants which only depend on
site ${\vect{i}}$.  To be more specific, $c^x_{\vect{i}} = 1/4L_x$ when the
site ${\vect{i}}$ locates at the lower-left or upper-right corner of
a shaded plaquette and $c^x_{\vect{i}} = -1/4L_x$ otherwise.  (Which
is `left' is an irrelevant question here as it is a matter
of the overall sign of the winding number and we are
interested only in its squared value).

Here, we should notice that the winding number $W_x^{(l)}$
or $W_y^{(l)}$ of a loop $l$ formed in the process of the
graph assignment can be expressed in terms of $c^x_{\vect{i}}$ and
$c^y_{\vect{i}}$ defined above as follows,
\begin{equation}
  W_x^{(l)} \equiv 2 \sum_{{\vect{i}}\in l} c^x_{\vect{i}} S_{\vect{i}}^z, \quad
  W_y^{(l)} \equiv 2 \sum_{{\vect{i}}\in l} c^y_{\vect{i}} S_{\vect{i}}^z.
\end{equation}
Then, the ordinary estimator ${\cal O}$ for the squared
winding number can be decomposed into two parts:
\begin{equation}
  {\cal O} = {\cal O}_{impr} + {\cal O}_{rem},
\end{equation}
where ${\cal O}_{impr}$ and ${\cal O}_{rem}$ are defined as
\begin{eqnarray}
  &&{\cal O}_{impr} = \frac14\left( \sum_l {W_{x}^{(l)}}^2 + \sum_l
    {W_{y}^{(l)}}^2 \right),\\ &&{\cal O}_{rem} = \frac14\left(
    \sum_{l \ne l'} W_{x}^{(l)} W_{x}^{(l')} +
    \sum_{l \ne l'} W_{y}^{(l)} W_{y}^{(l')}
  \right).
\end{eqnarray}

Since loops are flipped independently, the expectation
values of all the cross terms in ${\cal O}_{rem}$ are vanishing.
Therefore,
\begin{equation}
  \langle {\cal O}_{rem} \rangle = 0.
  \label{eq:rem}
\end{equation}
For the same reason, we can derive another useful equation,
\begin{equation}
  \langle {\cal O}_{impr} \cdot {\cal O}_{rem} \rangle = 0.
  \label{eq:imprRem}
\end{equation}
Equation (\ref{eq:rem}) leads to
\begin{equation}
  \langle {\cal O} \rangle = \langle {\cal O}_{impr}
  \rangle.
\end{equation}
This implies that the ${\cal O}_{impr}$ is another estimator
of the squared winding number.  The new estimator ${\cal
  O}_{impr}$ depends only on the graph variables.  We can
see that this new estimator is really ``improved'' as
follows.

Intuitively, one graph configuration represents $2^{N_l}$
spin configurations, where the $N_l$ is the number of loops
in the system. Therefore, a sampling of ${\cal O}_{impr}$
corresponds to taking an averaged value over many samplings
of ${\cal O}$.  From \eqref{imprRem}, the variance of ${\cal
  O}$ is related to those of ${\cal O}_{impr}$ and ${\cal
  O}_{rem}$ as
\begin{eqnarray}
  \mbox{Var}\left({\cal O}\right) &=& \langle({\cal O}_{impr}+{\cal
    O}_{rem})^2\rangle -\langle{\cal O}_{impr}+{\cal
    O}_{rem}\rangle^2\\ &=& \mbox{Var}\left({\cal O}_{impr}\right) +
  \mbox{Var}\left({\cal O}_{rem}\right) \\ &\ge& \mbox{Var}\left({\cal
      O}_{impr}\right).
  \label{eq:improved}
\end{eqnarray}
Equation (\ref{eq:improved}) is valid for other improved
estimators of the observable such as the one for the
susceptibility or structure factor.  Although we have
introduced the improved estimator based on the discrete time
representation, we can use the same definition in the
continuous time representation as well.

We compared the error between two estimators in long runs at
various temperatures.  We found that the new estimator
reduces errors in about twenty per cent and the total
performance of the new estimator is about $1.5$ times
better than that of the conventional one in terms of the
computational time required.

\subsection{Simulations}
\label{sec:XY-Sim}

In what follows, we present numerical results both from our
older set of simulations performed with a discrete time
version of the code and from our newer set with a continuous
time version.  We have taken various temperatures between
0.22 and 0.60 and used lattices with $L = 8, 12, 16, 24, 32,
48, 64, 96$ and $128$ in our simulation. For simulations
with the discrete time version, we have used $m=8,16$ and
$32$ for the Trotter number.  When the systematic error of
Trotter discretization exceeds the statistical error, we
have reduced the systematic error by the extrapolation to
$m=\infty$ using three different Trotter numbers.  For
$L=12,24,48,96$ and $128$ at all temperatures and for all
$L$'s near the critical temperature, we have used the
continuous time version.

The length of a typical run on $L=128$ at each temperature
is more than $10^6$ Monte Carlo sweeps~(MCS). The most
time-consuming part in the entire code is the cluster
identification.  It is a task of assigning each spin a
number that specifies which cluster the spin belongs to.  In
doing this, we only use the information of the local
connectivity.  To make a good use of vector processors for
this kind of task, we need to use a vectorizable algorithm.
For the discrete version of the code, we adopted an
efficient vectorized code following Mino's
idea\cite{Mino:91}.  This idea is based on the
``divide-and-conquer'' strategy. In this strategy, we
firstly divide the lattice into many small sub-lattices and
identifies clusters in each sub-lattice neglecting the
connectivity outside of the cluster. This process can be
easily vectorized or parallelized because cluster
identifications of different sub-lattices are independent of
each other. We then combine two adjacent original
sub-lattices to form a larger sub-lattice and identify
clusters in this sub-lattice.  In doing this, we can use the
information of the clusters in the smaller sub-lattices
which have already been obtained in the previous step.  We
repeat this procedure until all the sub-lattices are
combined into a single lattice, i.e., the original whole
lattice.  Using this algorithm, we achieved the efficiency
of 1.5 million site updates per second per one vector
processor on Fujitsu VPP500. For the loop algorithm on
continuous imaginary time, we used a parallel computer and
took trivial parallel approach.  Our code does about
$29000/(L^2\beta)$ sweeps per second on a node of Hitachi
SR2201.

In our simulations, each run is divided into several bins.
The length of a bin is taken large enough so that bin
averages may be statistically independent of each other at
least approximately. In order to check this condition, we
have measured autocorrelation times of the improved
estimator of the squared winding number at low temperatures
$T < 0.35J$ by the standard binning analysis for a run of
$10^5$ MCS with $L=64$. They turned out to be smaller than 2
MCS in all cases.  We have not observed any difficulty due
to low temperature except that, trivially, the computational
time per one Monte Carlo step increases proportional to the
inverse temperature.  The statistical independence among
bins is assured, because the smallest bin length used is
$1000$ MCS.  An error bar shown in the figures in the
present paper represents one standard deviation.  Results of
the squared winding number $\langle {\bf W}^2 \rangle$ are
summarized in \tabref{w2}.

\section{Universal Jump in the Helicity Modulus}
\label{sec:UJ-XY}

\subsection{Kosterlitz renormalization group equations}
\label{sec:XY-KTG}

The helicity modulus can be regarded as the renormalized
coupling constant that appears in the Kosterlitz
renormalization group equations.  Furthermore, Weber and
Minnhagen\cite{WeberM} regarded the renormalization group
flow of the solution as the system size dependence of the
helicity modulus.  They observed that the estimated values
at the critical temperature agree well with the theoretical
prediction derived from this idea.  In the present paper, as
we see below, we follow their idea but use a different
method for the analysis in which not only the data at the
critical temperature but also off-critical data are taken
into account simultaneously.

The Kosterlitz renormalization group equations are
\begin{equation}
  \frac{dx}{dl} = - y^2, \quad \frac{dy}{dl} = -
  xy. \label{eq:RG}
\end{equation}
Here, $x$ and $y$ are renormalized parameters after a
renormalization operation up to the length scale $L \equiv
L_0(T) e^l$ where $L_0(T)$ is some characteristic length of
the order of the lattice constant and has no singularity at
$T=T_{\rm KT}$.  The renormalized coupling constant $x$ is
related to the helicity modulus and the squared winding
number by the following equations\cite{NelsonK}:
\begin{equation}
  x = \frac{\pi\Upsilon}{T} - 2 = \frac{\pi}{2}\langle {\bf
    W}^2 \rangle - 2.
\end{equation}

Equation (\ref{eq:RG}) has an integral
\begin{equation}
  \Delta(T) \equiv x^2(l) - y^2(l),
\end{equation}
which does not depend on $l \equiv \log (L/L_0(T))$. As a
function of $T$, this integral has no singularity at
$T=T_{\rm KT}$.  Therefore we can expand it in terms of the
distance from the critical temperature, i.e., $\Delta(K)=a
(K-K_{\rm KT}) +b(K-K_{\rm KT})^2 + \cdots$ where $K \equiv
J/T$.  The solution of \eqref{RG} is given by
\begin{equation}
  x(T,L) = \left\{
    \begin{array}{cl}
      \sqrt{|\Delta|} \coth ( \sqrt{|\Delta|} l ) & \quad (K
      > K_{\rm KT}) \\ l^{-1} & \quad (K = K_{\rm KT})\\ 
      \sqrt{|\Delta|} \cot ( \sqrt{|\Delta|} l ) & \quad (K
      < K_{\rm KT}).
    \end{array} \right.\label{eq:Solution}
\end{equation}

This solution is a special case of the following form:
\begin{equation}
  x(T,L) = l^{-1} f(\Delta l^2). \label{eq:logFSS}
\end{equation}
This ``finite-size-scaling'' form can be obtained from the
ordinary one
\begin{equation}
  x(T,L) = L^{\omega} g\left( \Delta L^{1/\nu} \right )
  \label{eq:ordinaryFSS}
\end{equation}
by replacing $L$ by $l = \log(L/L_0(T))$.  However, it is
easy to see that \eqref{logFSS} cannot be made consistent with
\eqref{ordinaryFSS} no matter what values are used for the
exponents $\omega$ and $\nu$.  It should be also pointed out
that even if we consider a little more general form for the
ordinary finite size scaling
\begin{equation}
  x(T,L) = L^{\omega} g\left( \xi(T) / L \right)
\end{equation}
allowing $\xi(T)$ to have the correct temperature
dependence, it is still impossible to cast the solution
\eqref{Solution} into this form.

From the solution \eqref{Solution}, it can be seen that the
``scaling function'' $f(X)$ has no singularity at X=0. The
scaling function should take on the value 1 at the critical
point:
\begin{equation}
  f(X) = 1 + O(X). \label{eq:TheCriticalValue}
\end{equation}
Note also that our scaling form \eqref{logFSS} is
consistent with Olsson's fitting function [Equations (11a-c)
with (16) and (18) in ref.~\citen{Olsson}].

\subsection{Finite-size-scaling around $T_c$}
\label{sec:FSS-near-Tc}

In \figref{RawData}, the squared winding number is plotted
against the temperature.  We can see the strong system size
dependence characteristic to the KT transition especially
around and above the critical temperature.  Because of this
large size dependence it is virtually impossible to estimate
the critical temperature and the critical indices without
knowing the ``scaling form'' that describes the system size
dependence.  Figure \ref{fig:LogFSS} shows the same data
rescaled using \eqref{logFSS}.  The parameters $K_{\rm KT}$
and $L_0$ are chosen to minimize the cost function defined
in Appendix.

In \figref{Contour-8-12}, the contour plot of the cost
function is shown. The cost function is the measure of
the ``badness'' of the scaling plot.  It is basically the
deviation from the local linear approximation.  In order to
eliminate data out of critical region, we have to select
data points for the analysis.  We have eliminated data
points outside of the region,
\begin{eqnarray}
  & & \frac{4}{\pi} \le \langle {\bf W}^2 \rangle \le
  \frac{4}{\pi}+\frac12, \\ & & -1.5 \le x \equiv (K-K_{\rm
    KT}) (\log (L/L_0))^2 \le 1.5.
\end{eqnarray}
The value of the cost function at the optimal choice of the
parameters tends to be smaller as we eliminate more data
points away from the critical point, making the resulting
estimate more reliable. We have also selected data points 
with respect to the system size. However, if we eliminate too many
data points the cost function does not have a meaningful
minimum.  As long as we obtain a meaningful minimum, 
the results do not significantly depend on the
minimum system size adopted, as can be seen in
\figref{Contour-8-12}.  For the upper contour plot in
\figref{Contour-8-12} we have used data for system size
$L\ge 8$, whereas for the lower we used $L\ge 12$.

We here quote the estimates adopted in \figref{LogFSS}:
\begin{equation}
  T_{\rm KT} = 0.3427 (2)J,\quad L_0 = 0.29 (2).
  \label{eq:TKT}
\end{equation}
The minimum value of the cost function turns out to be 1.9.
The deviation of this value from unity suggests the presence
of correction to scaling. We can see from \figref{LogFSS},
that the value of the scaling function $f(x)$ at $x=0$ is
close to unity in agreement with the prediction
(\eqref{TheCriticalValue}).  The error in $f(0)$ was
determined in a similar fashion to those for $T_{\rm KT}$
and $L_0$.  It results in
\begin{equation}
  f(0) = 1.0(1).
\end{equation}
This agreement can hardly be
explained unless the Kosterlitz-Thouless picture is assumed
and is a clear indication for its validity in the present
model.

\subsection{Finite-size-scaling at $T_c$}
\label{sec:FSS-at-Tc}

We have also tried another analysis following Weber and
Minnhagen\cite{WeberM}, namely, measuring the chi-square
values of the fitting to the critical form 
(The second equation in \eqref{Solution})
as a function of the temperature.  To be more specific, we
assumed at each temperature the following system size
dependence of the helicity modulus.
\begin{equation}
  \frac{\pi\Upsilon}{2T} = \frac{\pi}{4} \langle {\bf W}^2
  \rangle = A(T) \left( 1 + \frac1{2 \log (L/L_0(T))}
  \right)
  \label{eq:FittingFunction}
\end{equation}
This fitting form is expected to be correct only at the critical
point with $A(T_{\rm KT})=1$ (\eqref{logFSS} and
\eqref{TheCriticalValue}).  

Since the number of data at each temperature is not enough,
the critical temperature cannot be determined as the one at
which the fitting is the best.  Instead, we have tried two
other procedures. One is to fix $A(T)$ to be 1 (but keep
$L_0$ as a fitting variable) and measure the chi-square
values of the fitting.  The result is shown in
\figref{chi2-WM-1}.  From this figure we conclude that
$T_{\rm KT} = 0.3430(5).$ The other procedure is to allow
the coefficient $A(T)$ to vary and see where $A(T)$ crosses
the line of $A=1$, which should be the case at the critical
temperature.  The estimated coefficient $A(T)$ as a function
of temperature while both $A$ and $L_0$ are allowed to vary
is shown in \figref{A}.  The critical temperature is
estimated by a linear fitting of the data, yielding
\begin{equation}
  T_{\rm KT} = 0.34271(5)J,
\end{equation}
which is consistent with \eqref{TKT}.

\section{Conclusion}
\label{sec:XY-Conclusion}
We have obtained accurate estimates of the helicity modulus
as a function of temperature and system size.  The results
fit into the special finite size scaling form derived from
the Kosterlitz renormalization group equation identifying
the renormalization scale with the system size.  By this
scheme we have avoided a technically difficult comparison
between an exponential divergence and an algebraic one.  The
coincidence of the estimated critical value of the scaling
function with the predicted one confirms the KT nature of
the phase transition.  In the numerical simulation we have
used loop algorithms in both discrete and continuous time
representations.  Both of them have turned out to be quite
efficient and advantageous especially in estimating the
helicity modulus which is usually a conserved quantity in
conventional Monte Carlo simulation.

A part of the numerical computation was done using Fujitsu
VPP500 at ISSP, the University of Tokyo and Kyoto University
Data Processing Center and Hitachi SR2201 at the computer
center of the University of Tokyo.  N.K.'s work is supported
by Grant-in-Aid for scientific research (No.~09740320) from
the Ministry of Education, Science and Culture of Japan.
%
%
\appendix
\section{Cost Function for Evaluating Finite-Size-Scaling Plots}
\label{app:cfss}

In this appendix, we present a new estimator for evaluating
finite-size-scaling plots.  The estimator is basically a
deviation from a local linear approximation analogous to the
previous estimator\cite{KawashimaI}.  The advantage of this
type of approaches is that we do not need to assume any
specific functional form for the scaling function. These
estimators are functions of trial values for various
parameters such as critical indices and temperature; $K_{\rm
  KT}$ and $L_0$ in the present case.  Since the previous
estimator was discontinuous as a function of these
parameters, searching for its minimum --- the optimal point ---
was a rather tricky task.  One could hardly tell if a local
minimum found was really a global minimum.  The
discontinuity came from the reordering of the data points as
we change the trial values of the exponents and the critical
temperature.  As we show in the following, we have
eliminated this discontinuity by treating data points for
different system size separately.

In what follows, a data point consists of a rescaled
estimate $y$, a rescaled error $d$ and a rescaled parameter
$x$ at which $y$ and $d$ are estimated. In the present case,
$y = ((\pi/2) \langle W^2 \rangle -2) \log (L/L_0)$ and $x =
(K-K_{\rm KT}) (\log (L/L_0))^2$.  Therefore, $x$, $y$ and
$d$ depend on trial values for various parameters in
general.  Here we denote one standard deviation by $d$.
Also we denote the $i$-th data point for the system size $L$
by $P_{L,i} \equiv (x_{L,i}, y_{L,i}, d_{L,i})$.  These data
points are ordered so that $x_{L,i} < x_{L,i+1}$.  In the
previous estimator, we dealt with all the data points
without regard to the system size. Therefore, the order of
the data points may change as the trial values for the
exponents vary.  Here we have no such reordering due to the
change in the trial values.

For each system size $L'$, we connect adjacent points
$P_{L',i}$ and $P_{L',i+1}$ by a straight line.  Then, for
each combination of a data point $P_{L,i}$ and a system size
$L'$, we choose $j$ such that $x_{L',j} < x_{L,i} <
x_{L',j+1}$.  Next, we consider a point at $x = x_{L,i}$ on
the line that connects two points, $P_{L',j}$ and
$P_{L',j+1}$.  We denote this point by $P_{L,i}^{L'}$.
Since the value $y_{L,i}^{L'}$ is expressed as a linear
combination of $y_{L',j}$ and $y_{L',j+1}$,
\begin{eqnarray}
  & & y_{L,i}^{L'} \equiv p y_{L',j} + q y_{L',j+1}, \\ & &
  p \equiv \frac{x_{L',j+1}-x_{L,i}}{x_{L',j+1} - x_{L',j}},\quad q \equiv 1-p,
\end{eqnarray}
the error $d_{L,i}^{L'}$, which comes from errors $d_{L',j}$
and $d_{L',j+1}$, can be estimated through the ordinary
error propagation rule.  To be specific,
\begin{equation}
  d_{L,i}^{L'} = \sqrt{ p^2 d_{L',j}^2 + q^2 d_{L',j+1}^2 }.
\end{equation}
For convenience, we define $P_{L,i}^{L} \equiv P_{L,i}$.  In
an ideal case where the linear approximation is good and
the statistical and systematic errors are negligible, all
$P_{L,i}^{L'}$'s for various $L'$ should coincide with each
other.

Here we assume that the linear approximation is good and the
systematic errors are negligible while the statistical
errors are not necessarily negligible.  In order to evaluate
the coincidence, then, we consider the weighted sum of
deviations of the points $P_{L,i}^{L'}$, i.e.,
\begin{equation}
  S_{L,i} \equiv \sum_{L'} \frac{1}{n-1}
  \left(\frac{y_{L,i}^{L'} - \bar
      y_{L,i}}{d_{L,i}^{L'}}\right)^2
\end{equation}
with
\begin{equation}
  \bar y_{L,i} \equiv \sum_{L'}
  \frac{y_{L,i}^{L'}}{{d_{L,i}^{L'}}^{2}} \left/ \sum_{L'}
    \frac1{{d_{L,i}^{L'}}^{2}} \right.,
\end{equation}
where $n$ is the number of distinct system sizes.  As
statistics tells, the expectation value of $S_{L,i}$ is $1$.
When we define $S$ as the sum of all $S_{L,i}$'s, it
evaluates quality of the finite-size-scaling plot.  Its
expectation value equals the number of data points $N$. The
acceptable range of parameters is determined by
\begin{equation}
  S < \min{S} + m \label{eq:ErrorCriterion},
\end{equation}
where $m$ is of order O(1).
%
%

\newpage
%
%
\begin{table}[htbp]
  \caption{The $v(G_p)$ and $a(G_p)$ for $S=1/2$ quantum $XY$ model.}
    \label{tab:v-ed-xy}
    \begin{tabular}{ccc}
      \hline
      $G_p$ & $v(G_p)$ & $a(G_p)$ \\
      \hline 1 & $\frac{1}{2}(e^{-\frac{\Delta\tau}{2}J} + 1)$
      & $-\frac{1}{4}J$\\
      2 & $\frac{1}{2}(e^{\frac{\Delta\tau}{2}J}-1)$ & $\frac{1}{4}J$\\
      3 & $\frac{1}{2}(1 - e^{-\frac{\Delta\tau}{2}J})$ & $\frac{1}{4}J$\\
      \hline
    \end{tabular}
\end{table}

\newpage
\begin{table}[htbp]
\caption{The squared winding number $\langle {\bf W}^2 \rangle$.}
\label{tab:w2}
\begin{tabular}{ccccc}
  \hline
  $T/J$ & $L=8$ & $L=16$ & $L=32$ & $L=64$ \\ \hline 0.220 &
  & 2.429(3) & 2.433(6) & 2.41(2)\\ 0.240 & & 2.214(3) &
  2.225(5) & 2.22(2)\\ 0.260 & & 2.040(3) & 2.046(5) &
  2.05(2)\\ 0.280 & & 1.876(2) & 1.865(4) & 1.86(2)\\ 0.300
  & & 1.728(2) & 1.722(4) & 1.73(2)\\ 0.320 & & 1.592(2) & &
  1.57(1)\\ 0.325 & 1.582(1) & 1.557(2) & 1.540(3) &
  1.538(5)\\ 0.330 & 1.552(1) & 1.524(2) & 1.508(3) &
  1.493(3)\\ 0.332 & 1.542(1) & 1.510(2) & 1.492(3) &
  1.478(3)\\ 0.334 & 1.528(1) & 1.492(2) & 1.478(3) &
  1.471(3)\\ 0.335 & 1.521(1) & 1.489(2) & 1.471(2) &
  1.454(3)\\ 0.336 & 1.518(1) & 1.483(1) & 1.463(2) &
  1.450(2)\\ 0.337 & 1.508(1) & 1.475(1) & 1.455(2) &
  1.446(2)\\ 0.338 & 1.505(1) & 1.466(1) & 1.444(2) &
  1.433(2)\\ 0.339 & 1.500(1) & 1.4617(9) & 1.442(1) &
  1.425(1)\\ 0.340 & 1.491(1) & 1.4556(9) & 1.434(1) &
  1.418(1)\\ 0.341 & 1.486(1) & 1.4480(9) & 1.426(1) &
  1.407(1)\\ 0.342 & 1.480(1) & 1.4417(9) & 1.419(1) &
  1.400(1)\\ 0.343 & 1.4727(9) & 1.4374(9) & 1.409(1) &
  1.392(1)\\ 0.344 & 1.4682(9) & 1.4292(8) & 1.403(1) &
  1.385(1)\\ 0.345 & 1.4633(9) & 1.4214(9) & 1.396(1) &
  1.377(1)\\ 0.346 & 1.4571(9) & 1.4149(9) & 1.388(1) &
  1.366(1)\\ 0.347 & 1.451(1) & 1.408(1) & 1.384(2) &
  1.359(2)\\ 0.348 & 1.445(1) & 1.401(1) & 1.370(2) &
  1.348(2)\\ 0.349 & 1.439(1) & 1.396(1) & 1.365(2) &
  1.342(2)\\

  0.350 & 1.433(1) & 1.391(1) & 1.356(2) &
  1.334(2)\\ 0.351 & 1.426(1) & 1.379(2) & 1.348(2) &
  1.319(2)\\ 0.352 & 1.423(1) & 1.376(1) & 1.338(2) &
  1.309(2)\\ 0.353 & 1.417(1) & 1.369(1) & 1.337(2) &
  1.308(2)\\ 0.354 & 1.412(1) & 1.360(1) & 1.327(2) &
  1.298(3)\\ 0.355 & 1.403(1) & 1.354(2) & 1.322(2) &
  1.286(3)\\ 0.356 & 1.395(2) & 1.349(2) & 1.311(2) &
  1.280(5)\\ 0.357 & 1.393(2) & 1.336(2) & 1.302(2) &
  1.270(5)\\ 0.358 & 1.386(2) & 1.336(2) & 1.298(2) &
  1.262(5)\\ 0.359 & 1.380(2) & 1.330(2) & 1.284(2) &
  1.255(5)\\ 0.360 & 1.373(1) & 1.322(2) & 1.279(2) &
  1.247(4)\\ 0.362 & 1.368(2) & 1.305(3) & 1.260(2) &
  1.21(1)\\ 0.364 & 1.356(2) & 1.296(2) & 1.246(2) &
  1.20(1)\\ 0.366 & 1.340(2) & 1.280(2) & 1.227(2) &
  1.18(2)\\ 0.368 & 1.328(2) & 1.262(2) & 1.207(2) &
  1.14(1)\\ 0.370 & 1.318(2) & 1.249(2) & & \\ 0.375 &
  1.290(2) & 1.214(2) & & \\ 0.380 & 1.259(1) & 1.180(2) & &
  \\ 0.390 & 1.203(1) & 1.105(2) & & \\ 0.400 & 1.146(1) &
  1.024(2) & 0.872(2) & 0.65(1)\\ 0.420 & 1.033(1) &
  0.856(2) & 0.598(2) & 0.242(6)\\ 0.440 & 0.917(1) &
  0.676(2) & 0.328(2) & 0.050(3)\\ 0.460 & 0.804(1) &
  0.492(2) & 0.142(1) & 0.007(1)\\ 0.480 & 0.686(1) &
  0.338(2) & 0.0518(7) & 0.0008(3)\\ 0.500 & 0.579(1) &
  0.214(1) & 0.0171(4) & \\ 0.520 & 0.479(1) & & & \\ 0.540
  & 0.388(1) & & & \\ 0.550 & & 0.0564(6) & & \\ 0.560 &
  0.310(1) & & & \\ 0.580 & 0.2465(9) & & & \\ 0.600 &
  0.1935(9) & 0.0132(3) & &\\
  \hline
\end{tabular}
\end{table}

\newpage
\setcounter{table}{1}
\begin{table}[htbp]
\caption{The squared winding number $\langle {\bf W}^2 \rangle$ (continued).}
\begin{tabular}{cccccc}
  \hline
  $T/J$ & $L=12$ & $L=24$ & $L=48$ & $L=96$ & $L=128$\\ 
  \hline 0.339 & 1.475(1) & 1.449(1) & 1.432(1) & 1.416(1) &
  1.416(1)\\ 0.340 & 1.468(1) & 1.442(1) & 1.423(1) &
  1.411(1) & 1.409(1)\\ 0.341 & 1.462(1) & 1.432(1) &
  1.414(1) & 1.401(1) & 1.398(1)\\ 0.342 & 1.4557(9) &
  1.426(1) & 1.409(1) & 1.392(1) & 1.389(1)\\ 0.343 &
  1.449(1) & 1.417(1) & 1.397(1) & 1.382(1) & 1.379(1)\\ 
  0.344 & 1.443(1) & 1.412(1) & 1.392(1) & 1.375(1) &
  1.370(1)\\ 0.345 & 1.4368(9) & 1.404(1) & 1.383(1) &
  1.366(1) & 1.363(1)\\ 0.346 & 1.4295(9) & 1.398(1) &
  1.374(1) & 1.357(1) & 1.350(1)\\ 
  \hline
\end{tabular}
\end{table}

\newpage

%
%
\deffig{breakups-xxz}{breakups.eps} { Three types of graphs
  for $S=1/2$ quantum $XY$ model. }

\deffig{RawData}{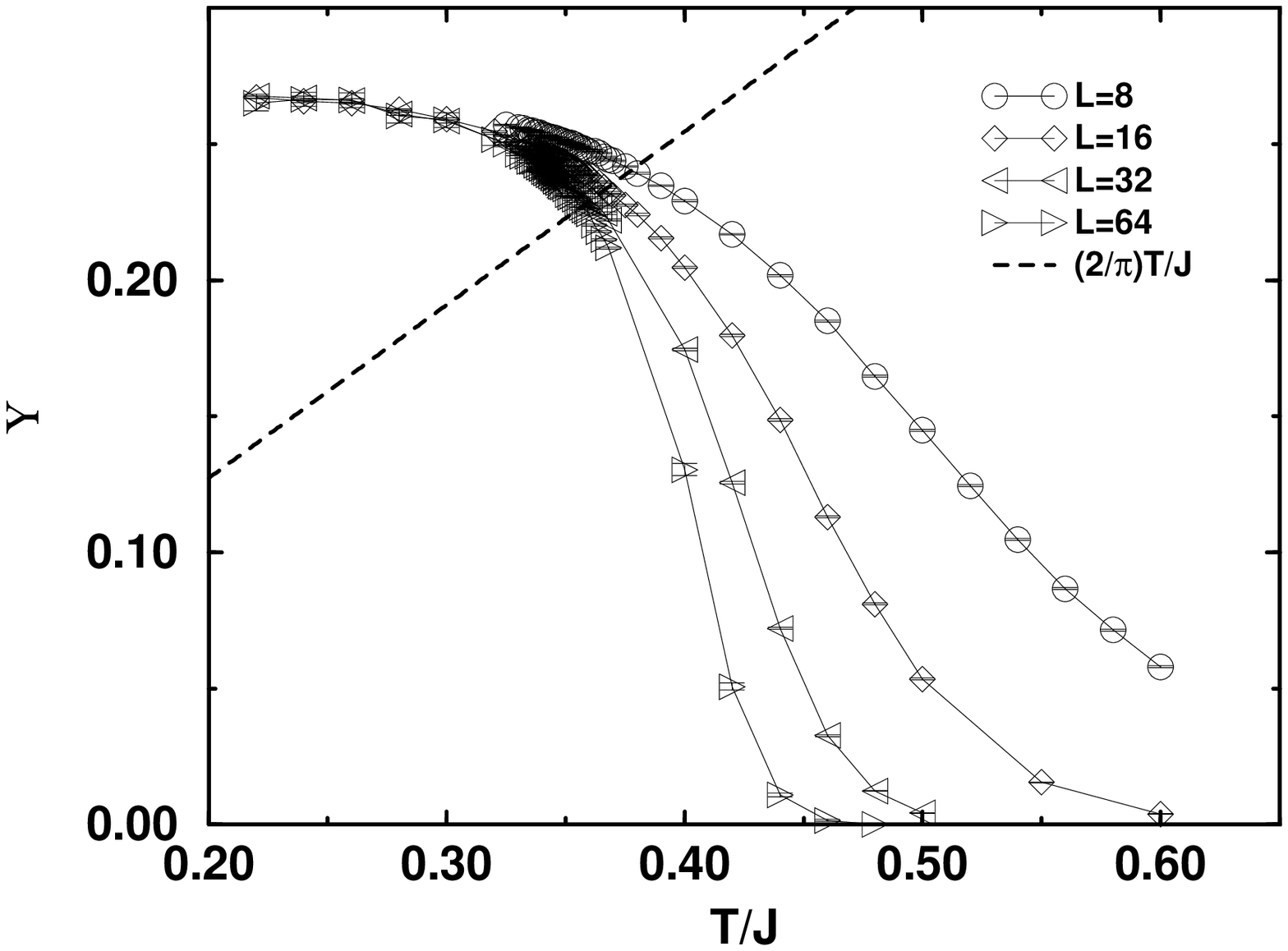} { Helicity modulus (or super fluid
  density) $\Upsilon = (T/2) \langle {\bf W}^2 \rangle$ as a
  function of temperature.  The universal jump is expected
  at the point where $\Upsilon = 2T/\pi$.  Error bars are
  drown but most of them are so small that they cannot be
  recognized.   }

\deffig{LogFSS}{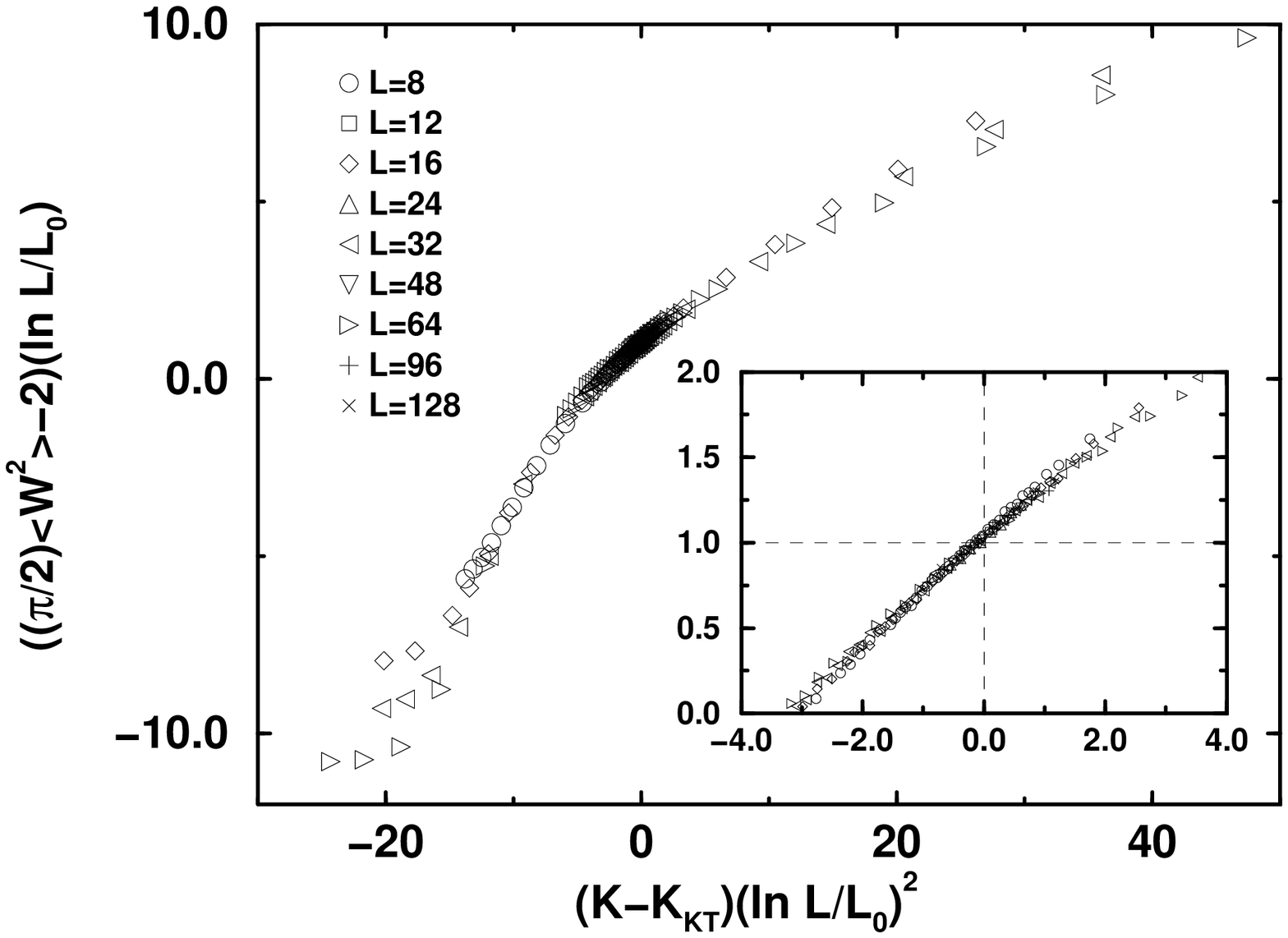} { A rescaled plot of the
  winding number fluctuation.  The inset is a enlarged view
  near the critical point.  }

\deffig{Contour-8-12}{fss-Lmin=8-12.eps} { Contour plots of
  the cost function for evaluating finite-size-scaling
  plots. The plotted value is defined as $S/N$ in Appendix.
  The smallest system size used is
  $L=8$ for the top figure and $L=12$ for the bottom.  }

\deffig{chi2-WM-1}{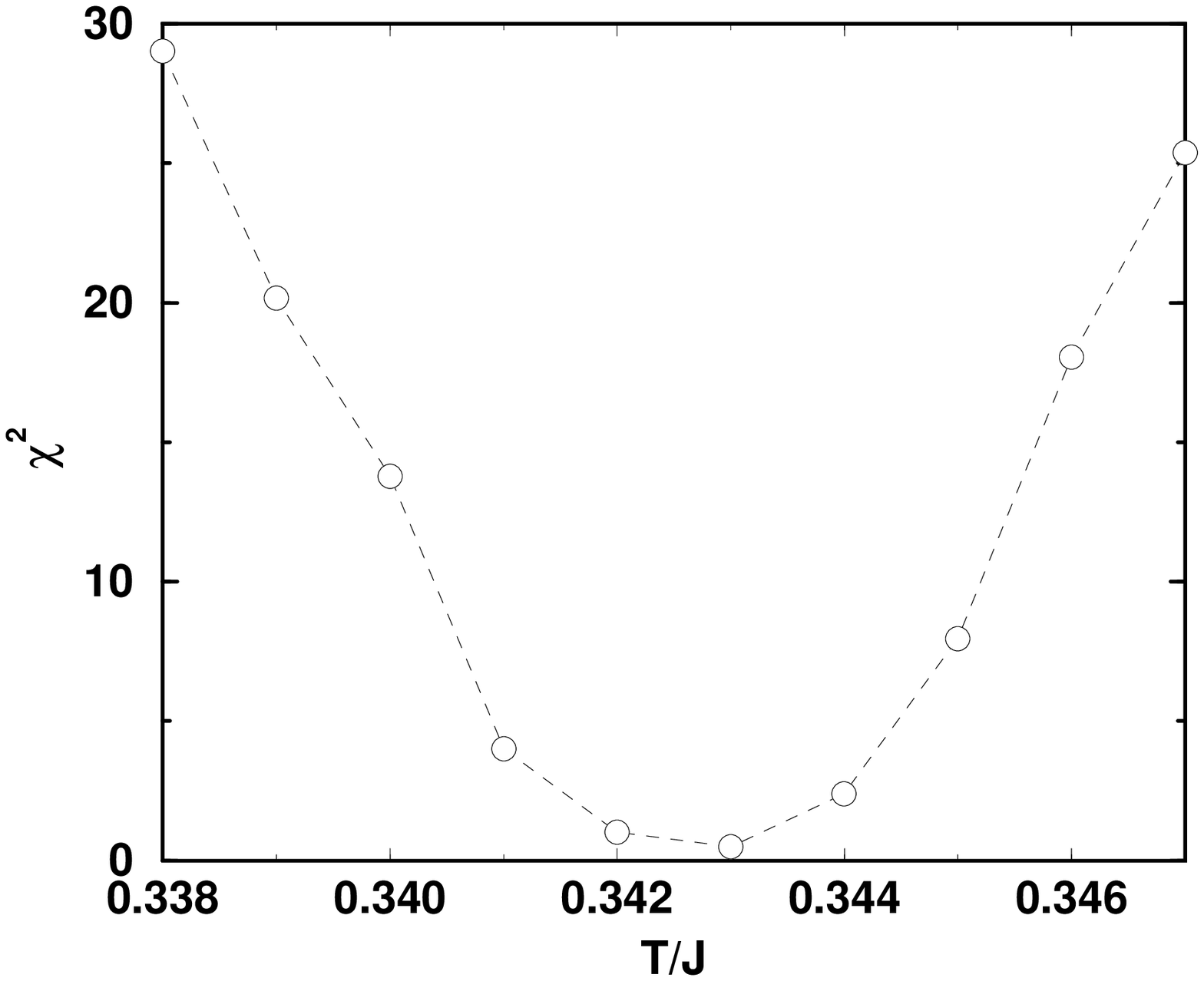} { Chi-square values of the
  fitting with $A(T)$ fixed to be 1.  }

\deffig{A}{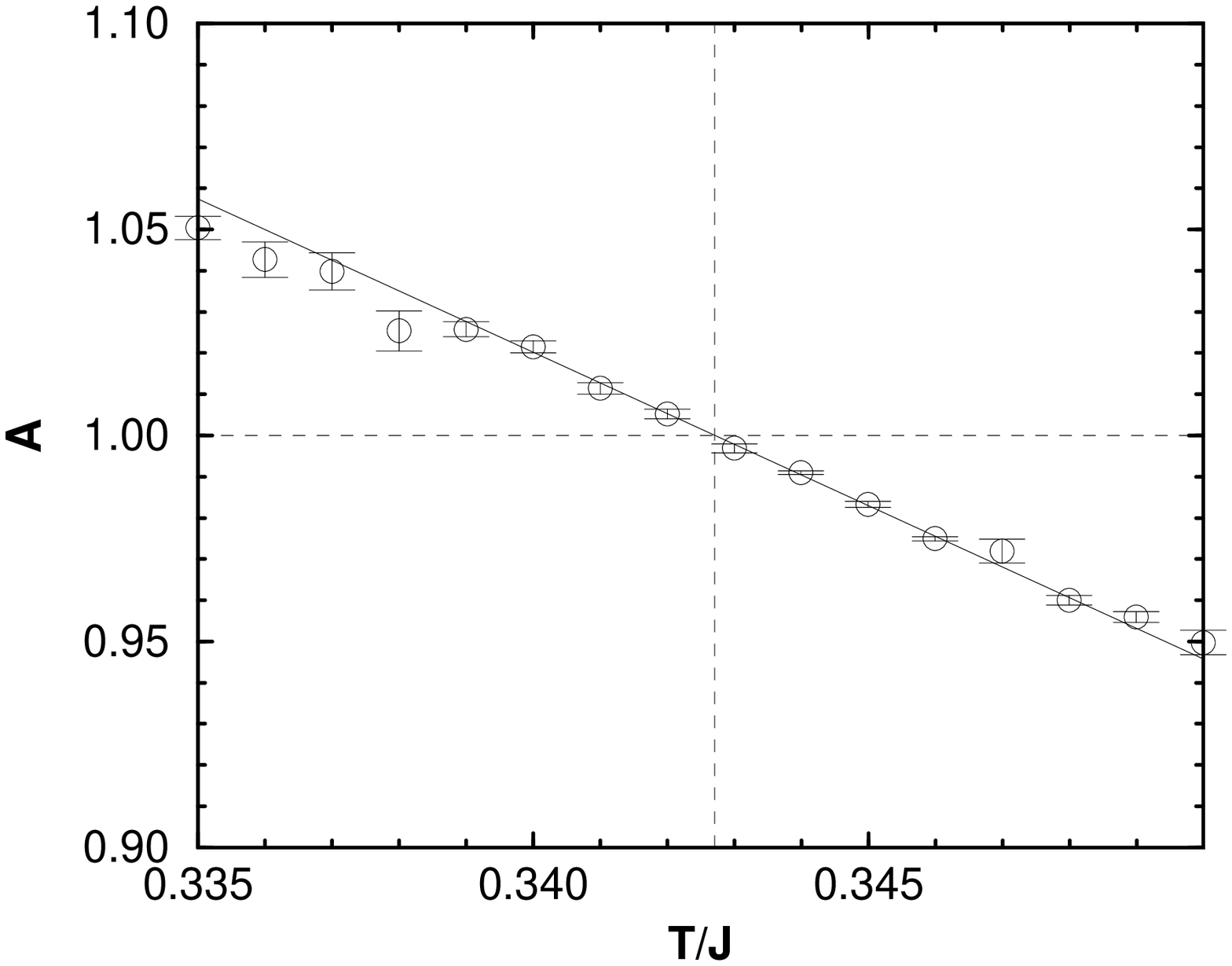} { Helicity modulus divided by the
  magnitude of its universal jump.  The dashed vertical
  line indicates the critical temperature at which the solid
  line crosses the dashed horizontal line ($A=1$).  The
  solid line is determined by a linear fitting.  }

\end{document}